\documentclass[11pt]{article}
\usepackage{graphicx}
\usepackage{lineno}
\usepackage{titlesec}
\usepackage{caption}
\usepackage{subcaption}
\usepackage{url}
\usepackage{comment}

\usepackage{setspace}
\let\oldbibliography\thebibliography
\renewcommand{\thebibliography}[1]{%
  \oldbibliography{#1}%
  \setlength{\itemsep}{0pt}%
}
\usepackage[margin=1.5in]{geometry}

\newcommand\pubdate{\today}

\def\Title#1{\begin{center} {\huge \bfseries #1 } \end{center}}
\def\Author#1{\begin{center}{ \sc #1} \end{center}}
\def\Address#1{\begin{center}{  #1} \end{center}}

\newcommand\pubblock{\rightline{\begin{tabular}{l}  \\ 
         \pubdate  \end{tabular}}}
\newenvironment{Abstract}{\begin{quotation}  }{\end{quotation}}
\newenvironment{Presented}{\begin{quotation} \begin{center} 
             PRESENTED AT\end{center}\bigskip 
      \begin{center}\begin{large}}{\end{large}\end{center} \end{quotation}}

\begin{document}
\begin{titlepage}
 \pubblock
\vfill
\Title{BSM physics using photon-photon fusion\\ processes in UPC in Pb+Pb collisions\\ with the ATLAS detector}
\vfill
\Author{Klaudia Maj\footnote{On behalf of the ATLAS Collaboration}}
\Address{AGH University of Kraków\\ 
al. Mickiewicza 30, 30-059 Kraków, Poland\\
\fontfamily{pcr}\selectfont 
e-mail: klaudia.maj@cern.ch}

\vfill
\begin{Abstract}
Relativistic heavy-ion beams at the LHC are accompanied by a large flux of equivalent photons, leading to multiple photon-induced processes. This proceeding presents searches for physics beyond the Standard Model enabled by photon-photon processes in both di-tau and diphoton final states. The tau-pair production measurements can constrain the tau lepton's anomalous magnetic dipole moment (g-2), and a recent ATLAS measurement using muonic decays of tau leptons in association with electrons and tracks provides one of the most stringent limits available to date. Similarly, light-by-light scattering proceeds via loop diagrams, which can contain particles not yet directly observed. Thus, high statistics measurements of light-by-light scattering provide a precise and unique opportunity to investigate extensions of the Standard Model, such as the presence of axion-like particles.
\end{Abstract}
\vfill
\begin{Presented}
DIS2023: XXX International Workshop on Deep-Inelastic Scattering and
Related Subjects, \\
Michigan State University, USA, 27-31 March 2023 \\
     \includegraphics[width=9cm]{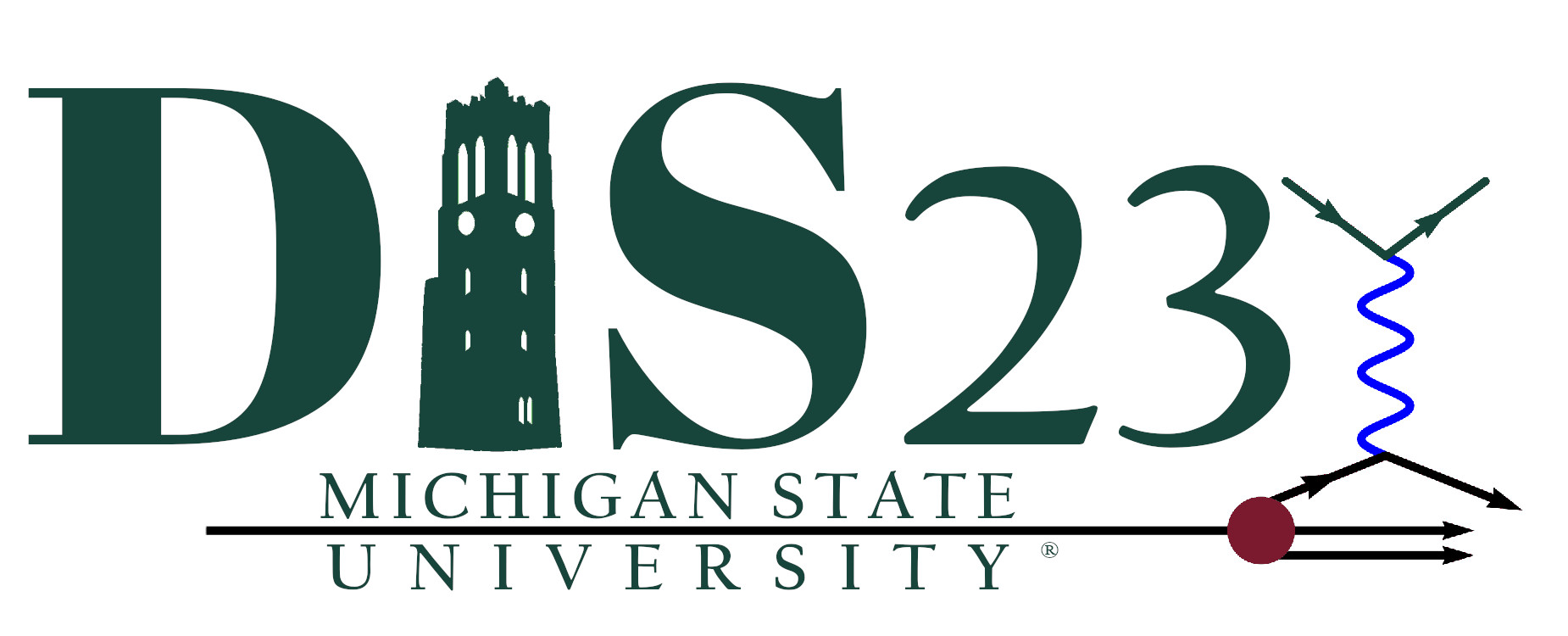}
\end{Presented}
\vfill
\end{titlepage}

\section{Introduction}
 In recent years, photon-induced processes in heavy-ion (HI) collisions have emerged as a promising path for studying beyond the Standard Model (BSM) physics. The ATLAS experiment \cite{ATLAS:2008xda} at the Large Hadron Collider (LHC) dedicates part of its annual operational time to the HI physics including
ultra-peripheral collisions (UPC).\ UPC are a unique category of HI collisions, which occur when the distance separating the interacting nuclei exceeds the sum of their radii. 
The large electromagnetic fields generated by relativistic ions can be considered as fluxes of photons - as described in the Equivalent Photon Approximation formalism \cite{Fermi:1925fq,PhysRev.45.729}.
Photon-photon interactions occur in both proton-proton, $pp$, and HI collisions. However, in the latter, the cross-sections for a specific process experience a significant increase due to the $Z^2$ scaling of the photon fluxes ($Z$ being the atomic number). Furthermore, HI collisions exhibit minimal hadronic pile-up, allowing for the identification of exclusive events and triggering on low-$p_{\mathrm{T}}$ particles.
This exceptional characteristics make UPC an excellent tool for studying rare processes and searching for BSM phenomena.
In this report two results of the ATLAS experiment \cite{ATLAS:2008xda} are discussed: the observation of $\gamma\gamma\rightarrow\tau^+\tau^-$ process with the measurement of anomalous magnetic moment of the $\tau-$lepton \cite{ATLAS:2022ryk} and the measurement of light-by-light (LbyL, $\gamma\gamma\rightarrow\gamma\gamma$) scattering with the search for axion-like particles (ALP) \cite{ATLAS:2020hii}. Both measurements utilize data from the UPC Pb+Pb collisions and are potentially sensitive to BSM effects.

\section{Exclusive $\tau^+\tau^-$ production and constraints on $a_{\tau}$}

ATLAS provides the exclusive observation of $\gamma\gamma\rightarrow\tau^+\tau^-$ process~\cite{ATLAS:2022ryk} using data from 2018 Pb+Pb collisions at $\sqrt{s_{\mathrm{NN}}}$= 5.02~TeV with an integrated luminosity of 1.44 nb$^{-1}$. The measurement of the exclusive production of $\tau$-leptons is used to set new constraints on the 
anomalous magnetic moment of the $\tau$-lepton, $a_{\tau}$. 
The theoretical SM prediction is: $a_{\tau}^{\mathrm{SM}} = 0.001 177 21 (5)$~\cite{Eidelman:2007sb}, which is remarkably
smaller than the currently available experimental bounds.
Various BSM theories, such as lepton compositeness, supersymmetry, 
and TeV-scale leptoquarks, etc., predicted modifications to the Standard Model (SM) 
value of $a_{\tau}$. The most stringent limits on $a_{\tau}$ are 
currently provided by the DELPHI experiment: 
$-0.052< a_{\tau} < 0.013$ 
(95\% CL)~\cite{DELPHI:2003nah}.

The identification techniques commonly employed in ATLAS cannot be used for signal $\tau$-leptons due to its very low transverse momentum ($p_{\mathrm{T}}$) values.\
Instead, it is required that events considered in the analysis contain one muon from $\tau$-lepton
decay, and electron or charged-particle track(s) from the
other $\tau$-lepton decay.
Three distinct signal regions (SR) are defined: muon and electron ($\mu e$-SR), muon and one track ($\mu$1T-SR), and muon and three tracks ($\mu$3T-SR). Candidate events are selected with a single muon trigger requiring muon $p_{\mathrm{T}}$ above 4 GeV. To ensure the exclusivity of the selected events a veto on forward neutron activity in the Zero Degree Calorimeter is imposed. 
Muons selected for the analysis are required to have $p_{\mathrm{T}}>$~4~GeV and $|\eta|<$ 2.4, selected electrons have $p_{\mathrm{T}}>$~4~GeV and $|\eta|<$ 2.47 and selected tracks should have $p_{\mathrm{T}}>$~100~MeV and $|\eta|<$ 2.5.  Events containing additional low-$p_{\mathrm{T}}$ tracks are rejected. Since different background processes contribute to each signal category, further requirements are introduced in the $\mu$1T-SR (muon and track system $p_{\mathrm{T}}>$ 1 GeV) and the $\mu$3T-SR (mass of the three-track system below 1.7 GeV). 
The main sources of background contributions arise from the exclusive dimuon production with the final-state radiation (FSR) and diffractive photonuclear interactions. The $\gamma\gamma\rightarrow\mu\mu$ background is constrained with a dimuon control region, 2$\mu$-CR. It requires exactly two opposite-charge muons with invariant mass above 11 GeV to suppress quarkonia backgrounds and no additional tracks separated from the muons.
\begin{figure}[!t]
    \centering
    \includegraphics[width=0.43\textwidth]{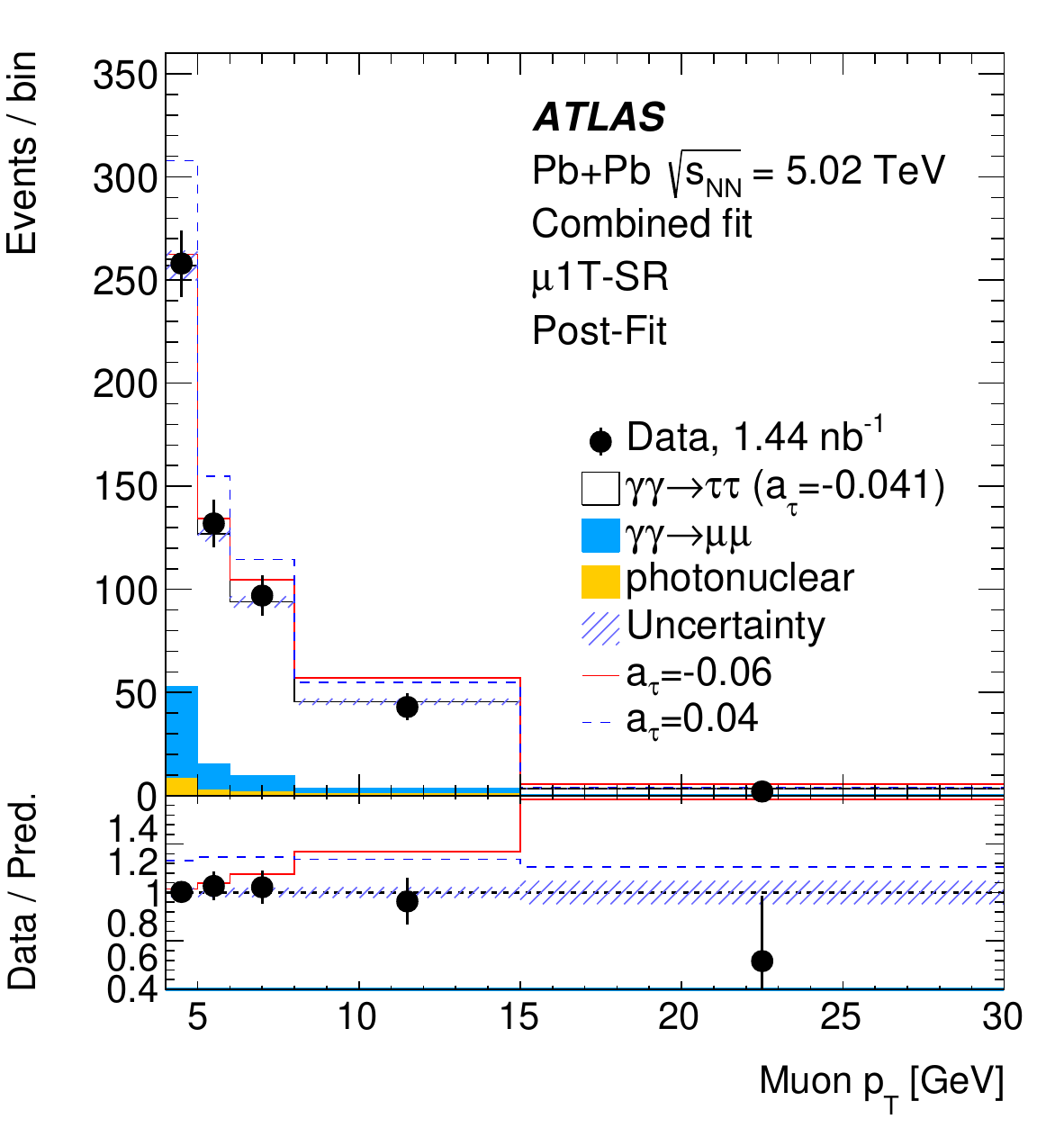}
    \hfill
    \includegraphics[width=0.43\textwidth]{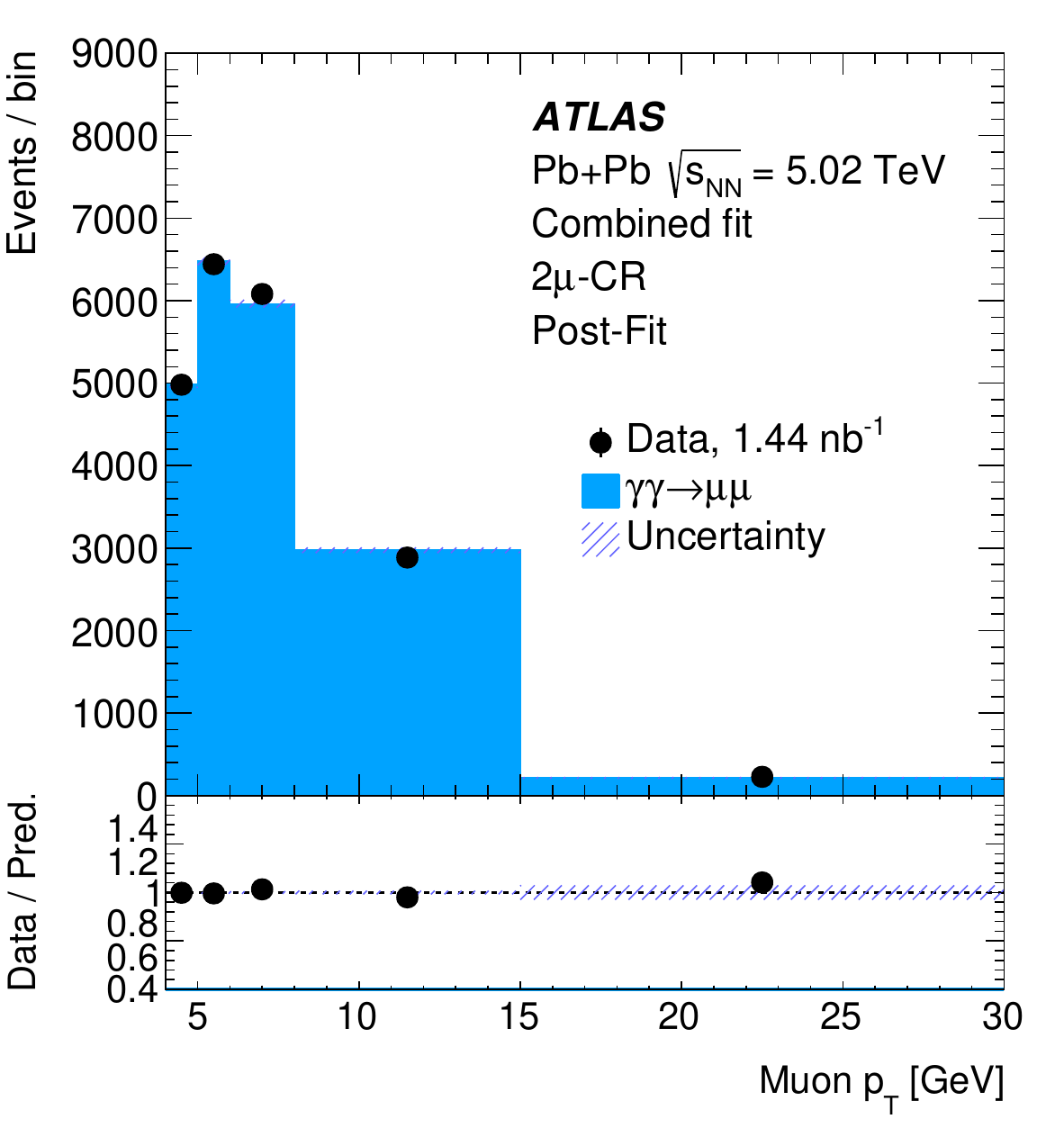}
    \caption{ Muon $p_{\mathrm{T}}$ distributions in the (left) 
    $\mu$1T-SR and (right) 2$\mu$-CR~\cite{ATLAS:2022ryk}.\ Post-fit distributions 
    are shown with the signal contribution corresponding to the 
    best-fit $a_{\tau}$ value ($a_{\tau}$=-0.042). For comparison, signal contributions with alternative $a_{\tau}$ values are shown.\ 
    }
    \label{fig:tau_dist}
\end{figure}

\begin{figure}[!h]
    \centering
    \includegraphics[width=0.53\textwidth]{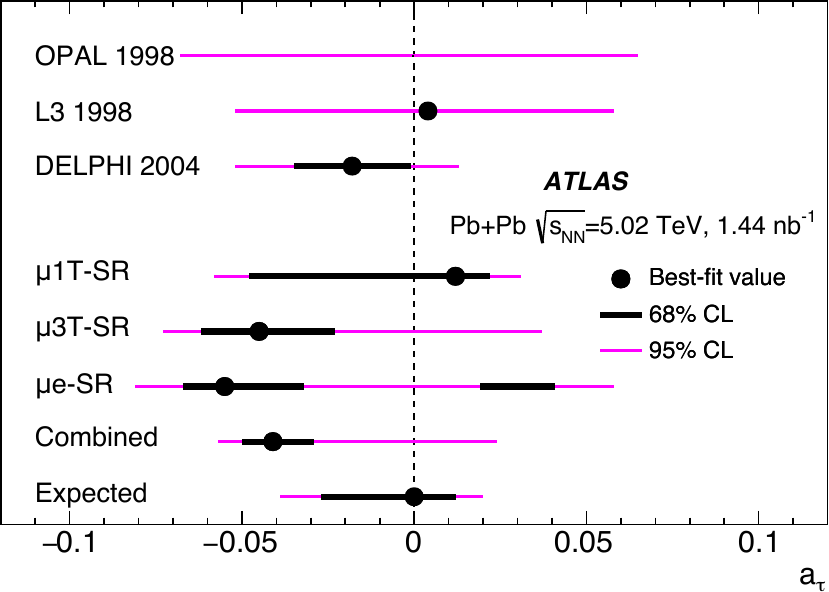}
         
     \caption{Measurements of $a_{\tau}$ from fits to individual signal regions (including the $2\mu$-CR), and from the combined fit~\cite{ATLAS:2022ryk}.\ These are compared with existing measurements from the OPAL, L3 and DELPHI experiments at LEP.\  }
      \label{fig:atau}
\end{figure}

After applying the event selection, a total of 656 data events were observed in three signal regions in which the analysis was performed.\ 
The fitted muon $p_{\mathrm{T}}$ distributions for the $\mu$1T-SR and $2\mu$-CR are shown in Figure~\ref{fig:tau_dist}. A very good data-to-prediction agreement is seen for the best-fit value of the $a_{\tau}$.
The $\gamma\gamma\rightarrow\tau^+\tau^-$ process was observed with a significance exceeding 5 standard deviations, and a signal strength of $\mu_{\tau\tau}$ = 1.03$_{-0.05}^{+0.06}$ assuming the SM value of $a_{\tau}$.\ 
To measure $a_{\tau}$, a fit to the muon $p_\mathrm{T}$ distribution is performed in the three SRs with $a_{\tau}$ being the parameter of interest. Also a control region with events from the $\gamma\gamma\rightarrow\mu^+\mu^-$ process is used in the fit to constrain initial-photon fluxes. 
Figure~\ref{fig:atau} presents a comparison of the ATLAS measurement of the anomalous magnetic moment of $\tau$-lepton in comparison to previous results obtained at the LEP experiments.\ The precision of this measurement is similar to the most precise single-experiment measurement by the DELPHI Collaboration~\cite{DELPHI:2003nah}.

\section{Light-by-light scattering and search for axion-like particles}

The $\gamma\gamma\rightarrow\gamma\gamma$ scattering is a rare phenomenon allowed by the quantum electrodynamics (QED) at the lowest order via a quantum loop of virtual charged  fermions or $W^{\pm}$ bosons.
LbyL production can be altered by various BSM phenomena: new particles entering the loop, Born-Infeld extensions of the QED, space-time non-commutativity in the QED, extra spatial dimensions, etc.
Furthermore, the diphoton mass spectrum obtained from the LbyL process can be explored to search for potential neutral axion-like particles, ALP. ALP may contribute to the distribution as a
narrow diphoton resonance~\cite{PhysRevLett.118.171801}.

LbyL scattering was also measured by ATLAS in UPC Pb+Pb collisions at $\sqrt{s_{\mathrm{NN}}} = $ 5.02 TeV using a combined 2015+2018 data sample with an integrated luminosity of 2.2~nb$^{-1}$~\cite{ATLAS:2020hii}.
The signature of interest is the exclusive production of two photons, each with transverse energy $E_{\mathrm{T}}^{\gamma} >$ 2.5 GeV, pseudorapidity $|\eta^{\gamma}| < $ 2.4
and diphoton invariant mass $m_{\gamma\gamma} >$~5~GeV with
transverse momentum $p_{\mathrm{T}}^{\gamma\gamma} <$ 1 GeV. 
Any extra activity in the detector is vetoed, in particular no reconstructed tracks originating from the
nominal interaction point with $p_{\mathrm{T}}>$~100~MeV are accepted.\ 
The final state photons are aligned in the azimuthal angle $\phi$. Back-to-back topology is studied using diphoton acoplanarity defined as $A_{\phi} = 1 - \frac{|\Delta\phi|}{\pi}$ . Event candidates
are expected to have $A_{\phi}< $ 0.01.\ 
The main background contribution originates from exclusive production of the electron–positron pairs ($\gamma\gamma\rightarrow e^+ e^-$). In the measurement, the $\gamma\gamma\rightarrow e^+ e^-$ background is suppressed with the requirement of no tracks and pixel-tracks reconstructed in the Inner Detector. A remaining dielectron contribution is evaluated using a data-driven method.
The second significant background source is gluon-induced central exclusive production (CEP) of photon pairs. The CEP background is evaluated using a dedicated control region in data ($A_{\phi}>$  0.01) and then extrapolated to the LbyL signal region.

ATLAS established the observation of LbyL process with 97 events observed in data, while the signal and background expectations are 45 events and 27 $\pm$ 5 events, respectively.
 The integrated cross-section measured in the fiducial phase space, defined by the requirements reflecting
 the event selection, is $\sigma_{fid} = 120 \pm 17(stat.)\pm13(syst.)\pm4(lumi)$ nb.\
The presented value can be compared with two theoretical predictions considered to be $78 \pm 8$ nb from the SuperChic v3.0 MC generator \cite{Harland-Lang:2018iur} and $80 \pm 8$ nb from \cite{PhysRevC.93.044907}. In addition to the integrated fiducial cross-section, ATLAS measured $\gamma\gamma\rightarrow\gamma\gamma$ differential cross-sections involving four kinematic variables of the final-state photons. In general, a good agreement between the measurement and SM predictions is found.

\begin{figure}[!t]
    \centering
    \includegraphics[width=0.47\textwidth]{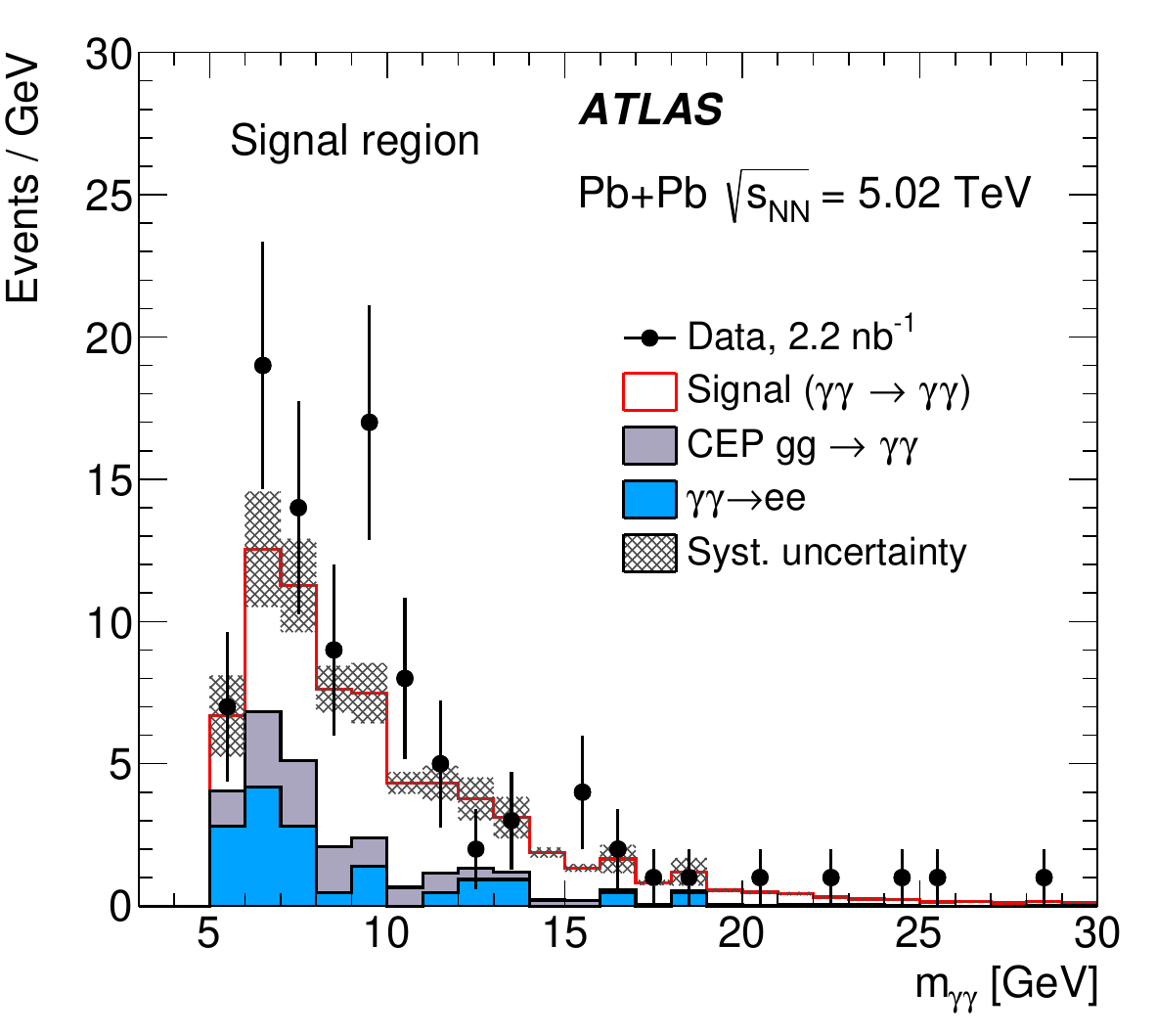}
     \caption{ The diphoton invariant mass for $\gamma\gamma\rightarrow \gamma\gamma$ event candidates~\cite{ATLAS:2020hii}. Data are compared with the sum of signal and background expectations.}
    \label{fig:lbyl1}
\end{figure}

\begin{figure}[!h]
    \centering
    \includegraphics[width=0.57\textwidth]{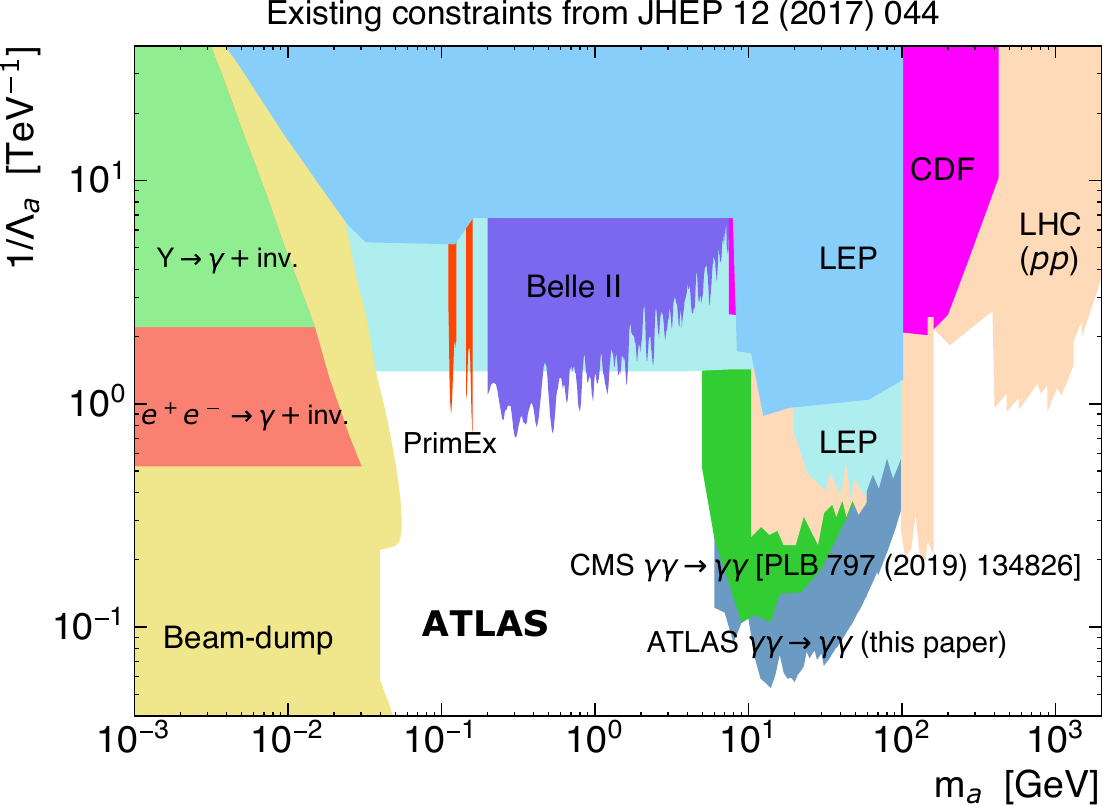}
     \caption{
     Compilation of exclusion limits at 95\% CL in the ALP-photon coupling (1/$\Lambda_{a}$) versus ALP mass ($m_{a}$) plane obtained by different experiments. The new limit obtained by ATLAS is marked with label: ATLAS $\gamma\gamma\rightarrow\gamma\gamma$~\cite{ATLAS:2020hii}}
    \label{fig:lbyl2}
\end{figure}

ALP may be produced in the photon–photon fusion, $\gamma\gamma\rightarrow a \rightarrow\gamma\gamma$, followed by the decay to the diphoton pair, where $a$ denotes the ALP field. Thus, a diphoton invariant mass distribution, $m_{\gamma\gamma}$, presented in Figure~\ref{fig:lbyl1}, can be interpreted for ALP searches. The ALP production would result in a resonance peak with diphoton mass equal to the mass of $a$.
The diphoton mass distribution was examined for
a mass range between 6 and 100 GeV. No significant excess of events over expected background was
found in the analysis. The 95\% CL
limit was derived for ALP production cross-section
and ALP coupling to photons $\frac{1}{\Lambda_{a}}$ as a function of ALP mass. A summary of exclusion limits from different experiments together with the new ATLAS constraints is shown in Figure~\ref{fig:lbyl2}. The new ATLAS analysis places the strongest limits on the ALP production in the intermediate mass region to date.

\section{Summary}
The report highlighted the significance of UPC in exploring rare SM processes and searching for BSM phenomena.\ The $\gamma\gamma\rightarrow\tau^+\tau^-$ process has been observed  in Pb+Pb UPC by the  ATLAS experiment, surpassing a 5$\sigma$ significance.\ The signal strength is consistent with the Standard Model expectations.\ The new constraints on the $a_{\tau}$ have been set, and are competitive to the best limits obtained during the LEP era.\ With the upcoming Run-3 data, an improvement in precision is anticipated.\ Additionally, the ATLAS experiment has established the presence of $\gamma\gamma\rightarrow\gamma\gamma$ scattering, with the results consistent
 with the Standard Model prediction.\ 
 The measured invariant mass of the diphoton system was used to set new exclusion limits on axion-like particles. This measurement provides the strongest constraints on the ALP production in the mass region of 6–100 GeV to date.

\section*{Acknowledgments}
This work was partly supported by program "Excellence initiative – research 
university" for the AGH University of
Kraków, by the National Science Centre 
of Poland under grant number UMO-2021/40/C/ST2/00187 and by PL-GRID infrastructure.\\
Copyright [2023] CERN for the benefit of the ATLAS Collaboration. CC-BY-4.0 license.

\bibliography{references}
\bibliographystyle{atlasBibStyleWoTitle}

\end{document}